\documentclass[12pt]{article}

\usepackage{newtxtext,newtxmath}

\usepackage{graphicx}

\usepackage[letterpaper,margin=1in]{geometry}

\renewenvironment{abstract}
	{\quotation}
	{\endquotation}

\date{}

\makeatletter
\renewcommand{\fnum@figure}{\textbf{Figure \thefigure}}
\renewcommand{\fnum@table}{\textbf{Table \thetable}}
\makeatother

\usepackage{scicite}

\usepackage{url}

\usepackage{siunitx}

\def\scititle{
	Turbulence at Low Reynolds Numbers
}
\title{\bfseries \boldmath \scititle}

\author{
	Ziyue~Yu$^{1}$,
	Xinyu~Si$^{1}$,
	Lei~Fang$^{1,2,3\dagger}$\and
    \small$^{1}$Department of Civil and Environmental Engineering, University of Pittsburgh, \and
    \small Pittsburgh, 15261, Pennsylvania, USA.\and
    \small$^{2}$Department of Mechanical Engineering and Materials Science, University of Pittsburgh, \and 
    \small Pittsburgh, 15261, Pennsylvania, USA.\and
    \small$^{3}$Department of Bioengineering, University of Pittsburgh, Pittsburgh, 15261, \and 
    \small Pennsylvania, USA.\\ [1ex]
	\small$^\ast$Corresponding author. Email: lei.fang@pitt.edu
}

\begin{document} 

\maketitle

\begin{abstract} \bfseries \boldmath
Turbulence—ubiquitous in nature and engineered systems—is traditionally viewed as an inherently inertial phenomenon, arising only when the Reynolds number, which quantifies the ratio of inertial to dissipative forces, greatly exceeds unity. Here, we demonstrate that strong energy transfer between length scales—a defining hallmark of turbulence—can persist even when the Reynolds number is near unity. We reinterpret this energy transfer as a mechanical interaction between turbulent stress and large-scale deformation. In Navier–Stokes turbulence, we introduce directionally biased perturbations that amplify this interaction, increasing energy flux by over two orders of magnitude in the absence of dominant inertia. Our findings establish a new regime of Navier–Stokes turbulence, challenging long-standing assumptions about the high Reynolds number conditions required for turbulent flows. Beyond revising classical views, our results offer a generalizable strategy for engineering multiscale transport in low Reynolds number environments such as microfluidic and biological systems.
\end{abstract}

\section{Introduction}
For over a century, it has been widely accepted that Navier–Stokes (N-S) turbulence—a chaotic, multiscale state of fluid motion—arises only when inertial forces exceedingly overwhelm energy dissipation \cite{kolmogorov1941local,kolmogorov1962refinement,kraichnan1967inertial,batchelor1969computation,pope2000turbulent}. This view is formalized through the Reynolds number (Re), a dimensionless quantity expressing the ratio of inertial to dissipation forces \cite{reynolds1883xxix,reynolds1895iv,orszag1973statistical}. Flows become unstable and turbulent when this quantity is much greater than one.
Underpinning this transition is the emergence of a spectral energy flux: energy transfers from injection scales toward the dissipation scale, which is eventually halted by dissipation mechanisms. In three-dimensional (3D) turbulence, this manifests as a forward cascade, where energy flows from large, energy-containing eddies to progressively smaller scales, ultimately dissipated by viscosity at the Kolmogorov scale \cite{kolmogorov1941local,kolmogorov1962refinement,pope2000turbulent}. In contrast, two-dimensional (2D) turbulence has an inverse energy flux where energy cascades from an injection scale to larger scales, which is eventually dissipated by friction \cite{kraichnan1967inertial,batchelor1969computation,frisch1995turbulence,xia2011upscale,boffetta2012two,fang2017multiple,fang2021spectral}. 

At low Re numbers, by contrast, flows are traditionally expected to remain smooth and reversible, governed by the linear balance between applied forcing and viscous or frictional damping \cite{purcell2014life,happel2012low}.
Here, we present a framework that challenges the classical assumption that turbulence is confined to high-Re regimes. We demonstrate that turbulence, characterized by strong spectral energy fluxes, can emerge even when the Re is on the order of unity. Notably, the magnitude of these fluxes rivals those seen in classical weak turbulence.

This low-Re turbulence arises from a mechanical reinterpretation of the spectral energy flux: the transfer of energy across scales can be viewed as mechanical work between stress (analogous to force) and rate of strain (analogous to displacement) at different scales. 
Under typical low-Re conditions, this stress-strain interaction is weak, leading to minimal energy flux and the appearance of smooth, laminar flow in physical space \cite{liao2013spatial}. 
Once recast in this mechanical view, it becomes evident that the geometric alignment between the stress and rate of strain tensors plays a critical role in determining the magnitude of energy transfer between different scales of motion.
We demonstrate that by geometrically aligning the principal directions of stress and rate of strain tensors, one can significantly increase spectral energy flux magnitudes even at Re near unity. 
Using electromagnetically driven thin-layer flows at Re on the order of unity, we introduce directionally biased physical perturbations to amplify spectral energy transfer by over two orders of magnitude, achieving spectral energy fluxes comparable to those observed in the traditionally defined weak-turbulence regime \cite{liao2013spatial,fang2016advection,fang2017multiple}. This approach enables us to generate sustained turbulence in a 2D N-S system under Re traditionally considered laminar. Our findings extend the known boundary of turbulent regimes and establish a path toward engineering turbulence in non-inertia-dominated environments, with applications ranging from microfluidics \cite{stroock2002chaotic,stone2004engineering} to biological systems \cite{katija2009viscosity,purcell2014life,houghton2018vertically,si2024biologically}. Extension of our framework to 3D is straightforward, with the primary difference being that 3D flows have three principal directions of stress and rate of strain, compared to two in 2D.

\section{Results}
\subsection{Theoretical framework of tensor alignment}
Filtering provides a foundational framework for analyzing scale interactions in nonlinear systems \cite{germano1992turbulence,liu1994properties} and has been applied across a wide range of physical systems \cite{rai2021scale,storer2022global,storer2023global,yang2023energy,zhang2023submesoscale,li2024eddy}. When a nonlinear equation is filtered at a characteristic length scale, the filtering process introduces additional terms that encapsulate the interactions between the retained and eliminated degrees of freedom. These terms effectively serve as sources or sinks for the dynamics of the retained scales. For example, applying a low-pass filter—which removes flow structures smaller than a cutoff scale \( L \)—to the N-S equations results in a subgrid-scale stress tensor, \(\tau_{ij}^{(\rm L)}\), that captures the momentum transfer across the scale \( L \) \cite{pope2000turbulent} (see Supplementary Material).

A similar approach to the filtered kinetic energy, defined as \({\rm E}^{(\rm L)} = \frac{1}{2}u_i^{(\rm L)}u_i^{(\rm L)}\), yields an expression for the spectral energy flux: \(\Pi^{(\rm L)} = - \tau_{ij}^{(\rm L)} s_{ij}^{(\rm L)}\), which quantifies the energy transfer between unresolved and resolved scales. Essentially, $\Pi^{(\rm L)}$ is the fundamental generator of turbulence, and significantly increasing its magnitude without substantially increasing Re is the goal of the paper. Here, \(u_i\) denotes the \(i^{\text{th}}\) component of the fluctuating velocity field and \(s_{ij}^{(\rm L)}\)
represents the filtered rate of strain tensor (see Supplementary Material). In this framework, \(\tau_{ij}^{(\rm L)}\) and \(s_{ij}^{(\rm L)}\) can be viewed analogously to force and displacement, respectively, such that their inner product reflects the work performed by the unresolved scales on the resolved ones through scale \(L\). 

This perspective underscores the critical role of the geometric alignment between the stress tensor \(\tau_{ij}^{(\rm L)}\) and the rate of strain tensor \(s_{ij}^{(\rm L)}\). When the two tensors are aligned, \(\Pi^{(\rm L)} < 0\), indicating an inverse energy cascade toward larger scales. In contrast, if the tensors are anti-aligned, \(\Pi^{(\rm L)} > 0\), signifying a forward cascade toward smaller scales. Furthermore, \(\Pi^{(\rm L)}\) can be recast as a function of the angular alignment between the eigenframes of \(\tau_{ij}^{(\rm L)}\) and \(s_{ij}^{(\rm L)}\). In two-dimensional flows, this relationship takes the form \cite{liao2014geometry,fang2016advection}:
\begin{equation} \label{eqn: tensor geometry}
    \Pi^{(\rm L)} = -2 \sigma \gamma  \cos(2\theta^{(\rm L)}),
\end{equation}
where \(\sigma\) and \(\gamma\) are the principal extensional eigenvalues of the rate of strain and the deviatoric part of the stress tensors, respectively, and \(\theta^{(\rm L)}\) is the angle between their corresponding extensional eigenvectors \(\hat{\sigma}\) and \(\hat{\gamma}\).

This formulation clearly shows that the orientation between \(\tau_{ij}^{(\rm L)}\) and \(s_{ij}^{(\rm L)}\) not only determines the direction of the energy flux but also modulates its magnitude. When \(\theta^{(\rm L)} < \pi/4\), the energy flux is directed toward larger scales, corresponding to an inverse cascade. Conversely, when \(\theta^{(\rm L)} > \pi/4\), energy is transferred toward smaller scales, indicating a forward cascade. No net energy flux occurs when \(\theta^{(\rm L)} = \pi/4\). In conventional isotropic 2D turbulence, the alignment between these tensors is governed by self-organization, typically resulting in a net inverse energy flux. Prior work has proposed using \(\eta = \cos(2\theta^{(\rm L)})\) as a quantitative measure of spectral flux efficiency, with experiments reporting a relatively low efficiency of about 27\% in isotropic 2D turbulence \cite{fang2016advection}. As the Re decreases, both the $\gamma$ and \(\theta^{(\rm L)}\) in Eqn. \ref{eqn: tensor geometry} decay rapidly to 0 and $\pi/4$, respectively, leading to laminar flow with only large-scale features and negligible spectral energy flux. 

In principle, a laminar flow with spatially varying velocity naturally exhibits a rate of strain tensor with non-negligible extensional eigenvalues. If a directionally biased perturbation, which can be represented spectrally as a stress tensor \(\tau_{ij}^{(\rm L)}\), is applied, then this externally imposed stress can couple with the existing rate of strain field in the low Re regime to produce a non-negligible spectral energy flux. Moreover, the geometric alignment between the stress and rate of strain tensors can also be enhanced by a deliberately imposed perturbation direction. 
Importantly, because the power of the physical perturbation remains small, the flow can still be kept at low Re regime.
As a result, it becomes possible to generate a low-Re turbulent flow that exhibits a much stronger energy flux, despite remaining in a nominally non-inertial-dominated regime. 

In our study, we selected hydrodynamic shear and cellular flows as the background configurations to establish a large-scale, well-aligned strain field (Fig. \ref{fig1}a and b), and applied a directionally biased monopolar forcing as a perturbation (Fig. \ref{fig1}e). The resulting direction of \(\hat{\gamma}\), associated with the monopolar perturbation, was found to align with the applied force direction (see Methods). By precisely tuning the mechanical angle \(\theta\) between the strain eigenvector \(\hat{\sigma}\) of the background flow and the direction of the monopolar force \textit{a priori}, we achieved more than two orders of magnitude enhancement in spectral energy flux at the same Re. The magnitude of the spectral energy flux achieved at a Re on the order of one is comparable to that traditionally observed in weak turbulence regimes \cite{liao2013spatial,fang2016advection}. 

Here, we define the Re as $\mathrm{Re_\alpha} = u / (L_i \alpha)$, where u is the root-mean-square velocity, $L_i$ is the characteristic length scale—chosen to be the spacing between rods (1.5 cm)—and $\alpha$ is the linear friction coefficient \cite{boffetta2012two}. The rod spacing is used because it represents the only obvious length scale in our experimental setup. For the rest of the paper, we will use Re for simplicity. In the quasi-2D flow studied here, energy dissipation is dominated not by viscosity but by linear friction with the bottom boundary. As a result, the Re defined in this context characterizes the ratio of inertial forcing to frictional dissipation. This formulation provides a more physically appropriate measure of momentum-dissipation balance in flows where conventional viscous effects are subdominant.

\subsection{Experimental methods} 

\begin{figure}[p]
    \centering
    \includegraphics[width = \textwidth]{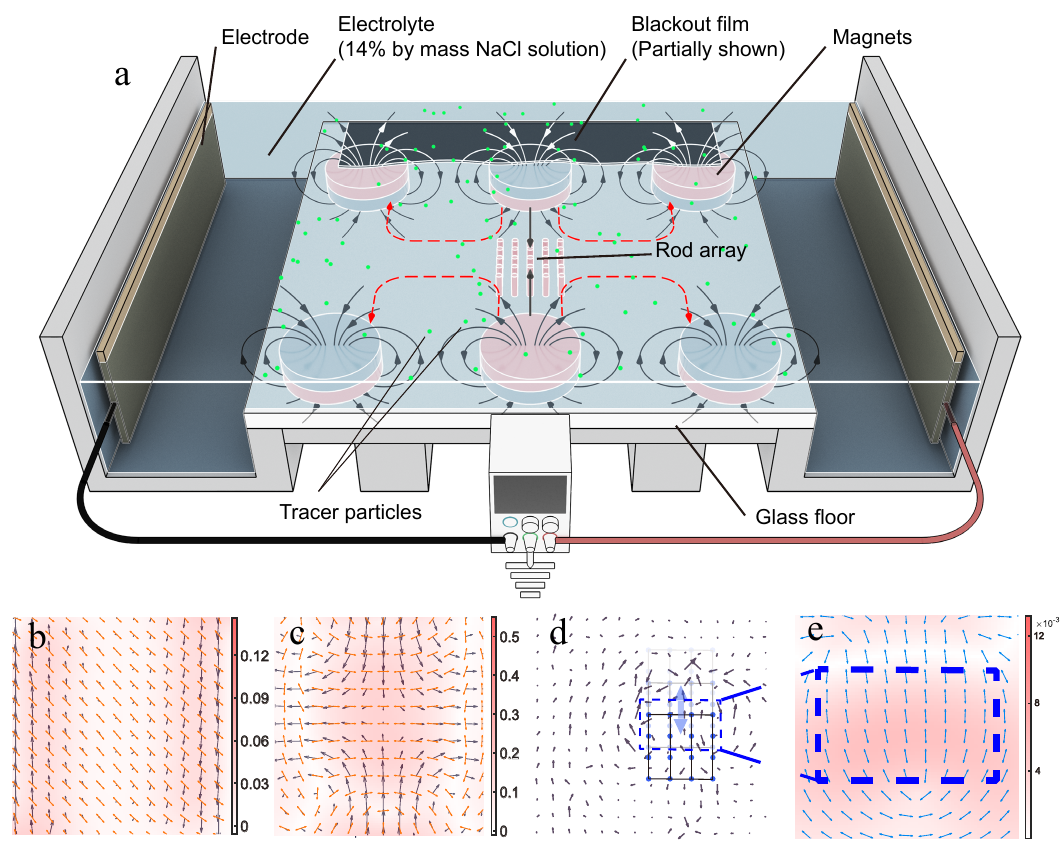}
    \caption{\textbf{Experimental setup and the characterization of the background flow and physical perturbations.} (\textbf{a}) A schematic of the experimental setup for cellular flow. The vertical magnetic field from the permanent magnets interacts with the horizontal direct current to generate the Lorentz force on the fluid, which acts nearly within the plane. The tracer particle on the fluid’s surface represents the 2D space under study. A rod array is controlled by a linear actuator. The generation of shear flow only requires the rearrangement of magnets (see Supplementary Material). (\textbf{b}) The flow field and the extensional direction of $s_{ij}^{(\rm L)}$ ($\hat{\sigma}$) of the hydrodynamic shear. (\textbf{c}) The flow field and $\hat{\sigma}$ of cellular flow. Red double-headed arrows indicate $\hat{\sigma}$, and the black arrows are the velocity field (down-sampled for clarity). (\textbf{d}) Flow field generated by rods moving in a quiescent fluid. (\textbf{e}) Field of extensional direction of $\tau_{ij}^{(\rm L)}$ ($\hat{\gamma}$). 
    }
    \label{fig1}
\end{figure}

To apply this theoretical framework for significantly enhancing spectral energy flux at Re on the order of unity, we conducted a series of controlled experiments using an electromagnetically driven thin-layer flow system \cite{fang2017multiple,fang2018influence,fang2019local,si2024biologically,si2025manipulating}. Two distinct base flow configurations were established—shear (Fig. \ref{fig1}b) and cellular flow (Fig. \ref{fig1}c)—each generated via Lorentz body forces arising from the interaction between a magnetic field produced by permanent magnets and a direct current across a thin electrolyte layer (Fig. \ref{fig1}a). Both flow types exhibited well-organized, large-scale rate of strain tensor structures (Fig. \ref{fig1}b and c). 

Controlled perturbations were introduced through periodic oscillation of a $4 \times 4$ rod array actuated by a programmable linear actuator.
For all the experiments, we make the rod move at around 2.5 times the background root-mean-square velocities. The extensional direction of $\tau_{i,j}^{(L)}$ is aligned with the moving direction (Fig. \ref{fig1}d and e), which is predicted by a theory (see Methods). The flow field, spanning $20 \times 20$\,cm$^2$, was recorded using an industrial camera and analyzed via particle tracking velocimetry (PTV) algorithm \cite{ouellette2006quantitative} (see Methods). 

\subsection{Turbulence at low Reynolds number} 
\begin{figure}[p]
    \centering
    \includegraphics[width = \textwidth]{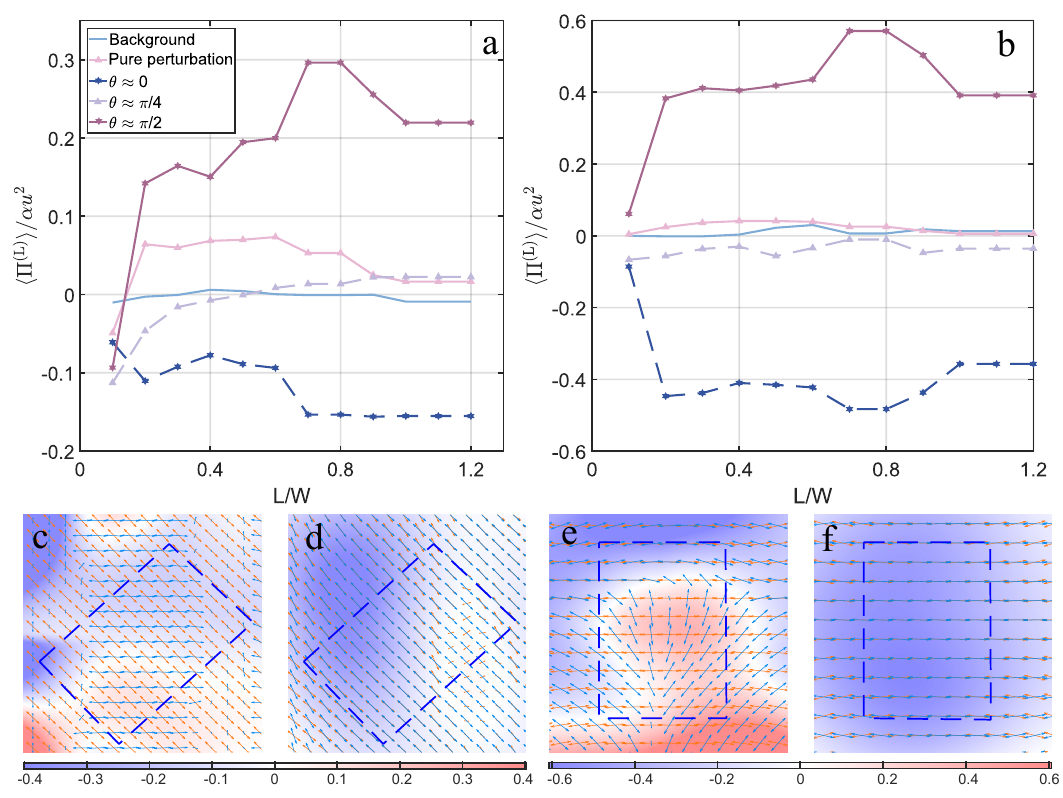}
    \caption{\textbf{Experimental results of energy flux enhancement.} This figure illustrates how the added directionally biased physical perturbation can significantly increase the energy flux by enhancing the tensor alignment.
    (\textbf{a} and \textbf{b}) The $\Pi^{(\rm L)}$ for different cutoff lengths for shear and cellular flows, respectively. (\textbf{c} and \textbf{e}) The tensor orientation without physical perturbation for shear and cellular flows, respectively. (\textbf{d} and \textbf{f}) The tensor orientation with physical perturbation for the case with mechanical angle $\theta \approx 0$ for shear and cellular flows, respectively. 
    Red double-headed arrows are the extensional direction of the local $s_{ij}^{(\rm L)}$ ($\hat{\sigma}$), and blue double-headed arrows are the extensional direction of the local $\tau_{ij}^{(\rm L)}$ ($\hat{\gamma}$). The background color represents the spatial distribution of instantaneous $\Pi^{(\rm L)}$. The results in \textbf{c}-\textbf{f} are based on a cutoff length scale of 0.7. The blue dashed box outlines the specific subdomain of the flow field used for all subsequent analysis presented in this paper. All lengths are normalized by half of the measurement domain size W. $\Pi^{(\rm L)}$ is normalized by the frictional dissipation $\alpha u^2$. 
    For \textbf{a}, $Re$ = 1.23 (background), 0.15 (pure perturbation), 1.27 ($\theta \approx 0$), 1.45 ($\theta \approx \pi/4$), and 1.38 ($\theta \approx \pi/2$). For \textbf{b}, $Re$ = 1.43 (background), 0.04 (pure perturbation), 1.14 ($\theta \approx 0$), 1.26 ($\theta \approx \pi/4$), and 1.12 ($\theta \approx \pi/2$). 
    For \textbf{c}-\textbf{f}, $Re$ = 1.23, 1.27, 1.43, and 1.14, respectively.
    }
    \label{fig2}
\end{figure}

In Fig. \ref{fig2}, we observed a significant energy flux enhancement in both shear and cellular flows. We had two control cases. The first is the background flow. We will contrast the energy flux magnitude of the background flow with that of the enhanced cases. In the second case, the rod array moved in a quiescent fluid to rule out the possibility that the enhanced energy flux is due to the motion of the rod array itself. Experiments were grouped based on the mechanical angle \(\theta\). 
One obvious observation one can make is that the direction of energy flux can be manipulated based on  \(\theta\) \textit{a priori} \cite{si2025manipulating}. In this paper, we will focus on the significant enhancement of the energy flux magnitude at low Re. At Re on the order of one, we observe that the energy flux magnitude is significantly enhanced in \(\theta \approx 0 \) and \(\theta \approx \pi/2 \) cases (Fig. \ref{fig2}a and b).

\begin{figure}[p]
    \centering
    \includegraphics[width = \textwidth]{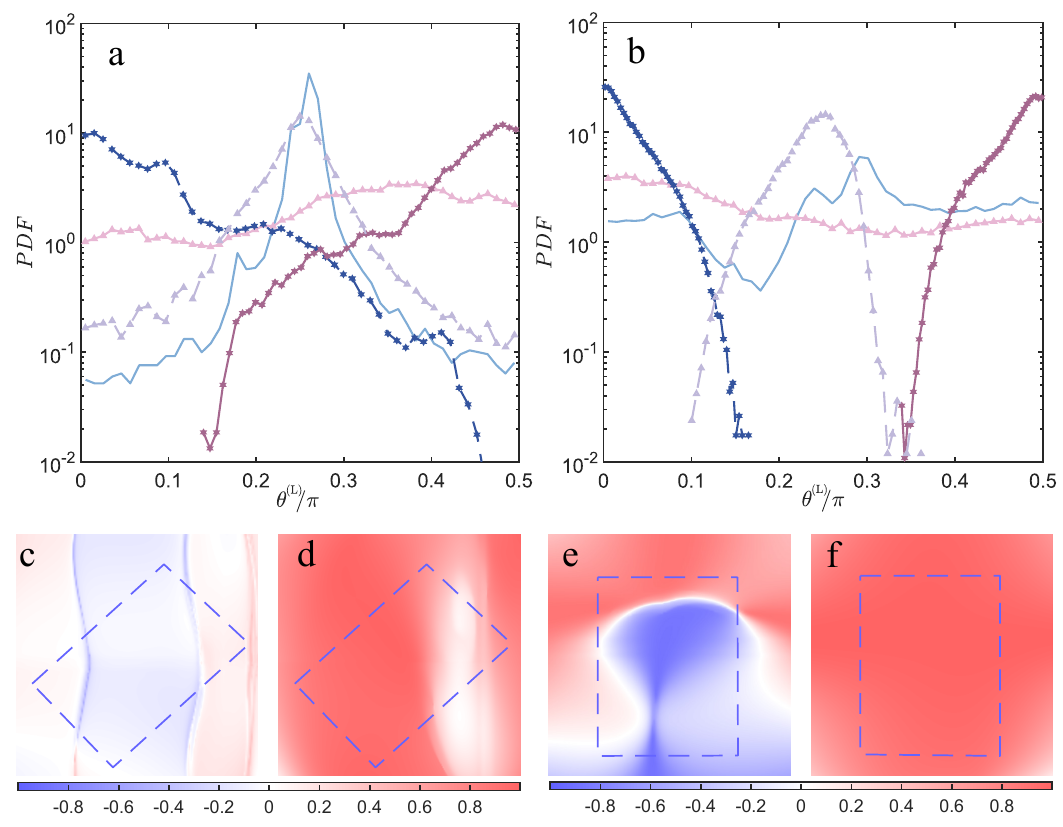}
    \caption{\textbf{Experimental results of stress-strain orientation statistics.} (\textbf{a} and \textbf{b}) Probability density functions of $\theta^{(L)}$ for shear and cellular flows, respectively. (\textbf{c} and \textbf{e}) Efficiency, $\eta$, without physical perturbation for shear and cellular flows, respectively. (\textbf{d} and \textbf{f}) Efficiency, $\eta$, with physical perturbation for hydrodynamic shear and cellular flows. The results in \textbf{c}-\textbf{f} are based on a cutoff length scale of 0.7. The blue dashed box outlines the specific subdomain of the flow field used for all subsequent analysis presented in this paper. All lengths are normalized by half the domain size W. $\Pi^{(\rm L)}$ is normalized by the frictional dissipation $\alpha u^2$. 
    Since the perturbed flows include additional energy input, the Re differs between perturbed and unperturbed cases. For \textbf{a}, $Re$ = 1.23 (background), 0.15 (pure perturbation), 1.27 ($\theta \approx 0$), 1.45 ($\theta \approx \pi/4$), and 1.38 ($\theta \approx \pi/2$). For \textbf{b}, $Re$ = 1.43 (background), 0.04 (pure perturbation), 1.14 ($\theta \approx 0$), 1.26 ($\theta \approx \pi/4$) and 1.12 ($\theta \approx \pi/2$). 
    For \textbf{c}-\textbf{f}, $Re$ = 1.23, 1.27, 1.43 and 1.14, respectively.
    The legend is the same as that of Fig. \ref{fig2}. }
    \label{fig3}
\end{figure}

Tensor geometry provides a clearer understanding of the energy flux magnitude enhancement. In Fig. \ref{fig2}c-f, we show the tensor alignment (\(\theta^{(L)}\)) between the extensional direction of stress and the rate of strain tensors. Without manipulation, shear (Fig. \ref{fig2}c) and cellular (Fig. \ref{fig2}e) flows have \(\theta^{(L)}\) mostly close to $\pi/4$ leading to very small energy flux magnitudes. With directionally biased physical perturbation at \(\theta \approx 0\), the perturbed shear (\ref{fig2}d) and cellular (\ref{fig2}f) flow have stress and rate of strain eigenframe well aligned ($\theta^{(L)} = 0$), causing an enhanced inverse energy flux and close-to-optimal $\eta$. In Fig. \ref{fig3}a and b, we observe a significant correlation between \(\theta\) and \(\theta^{(L)}\) even at Re on the order of unity. For background flows, pure perturbation and \(\theta \approx \pi/4 \) cases, we only observe a weak asymmetry around $\pi/4$ as compared to the \(\theta \approx 0 \) and \(\theta \approx \pi/2 \) cases. As a consequence, we see that the spatial distribution of $\eta$ is significantly enhanced for \(\theta \approx 0 \) cases for both shear and cellular flows. In Fig. \ref{fig3}c-f, we observe the spatial distribution of $\eta$. With the manipulation, we observe that the $\eta$ is significantly enhanced.  

Notably, the tensor geometry framework remains effective even at Re on the order of unity, where inertial forces are non-dominant. In such regimes, flows are conventionally classified as laminar, and persistent spectral energy flux is generally considered absent in such regimes. However, our results challenge this assumption: despite the low Re, we observe clear and non-negligible scale-to-scale energy transfer, driven by the alignment between the $s_{ij}^{(\rm L)}$ and $\tau_{ij}^{(\rm L)}$. This behavior is rooted in the N-S equations. The Re characterizes the relative strength of advection to dissipation—viscosity in three-dimensional flows, or linear friction in quasi-two-dimensional systems. When this ratio is finite, even if small, advection remains dynamically important. Consequently, the tensorial interactions responsible for energy flux—governed by stress–strain alignment—persist and remain meaningful. These observations indicate that the mechanisms underpinning spectral energy transport are not exclusive to high-Re turbulence but can extend into regimes previously deemed dynamically trivial. 

Under generic forcing conditions at low Re, where advection and dissipation are comparable, spectral energy transfer between scales remains negligible and, as a spatial manifestation, flows exhibit a laminar behavior. This limited flux arises from the geometric misalignment between $s_{ij}^{(\rm L)}$ and $\tau_{ij}^{(\rm L)}$. However, our results demonstrate that if perturbations are engineered to align these two tensors, sustained and appreciable energy transfer between scales can be induced, even when the $\mathrm{Re} \sim \mathcal{O}(1)$. This finding suggests that turbulence, traditionally associated with high-Re regimes, may be actively constructed and sustained in nominally laminar systems via geometric control of tensor alignment.

\begin{figure}[p]
    \centering
    \includegraphics[width = \textwidth]{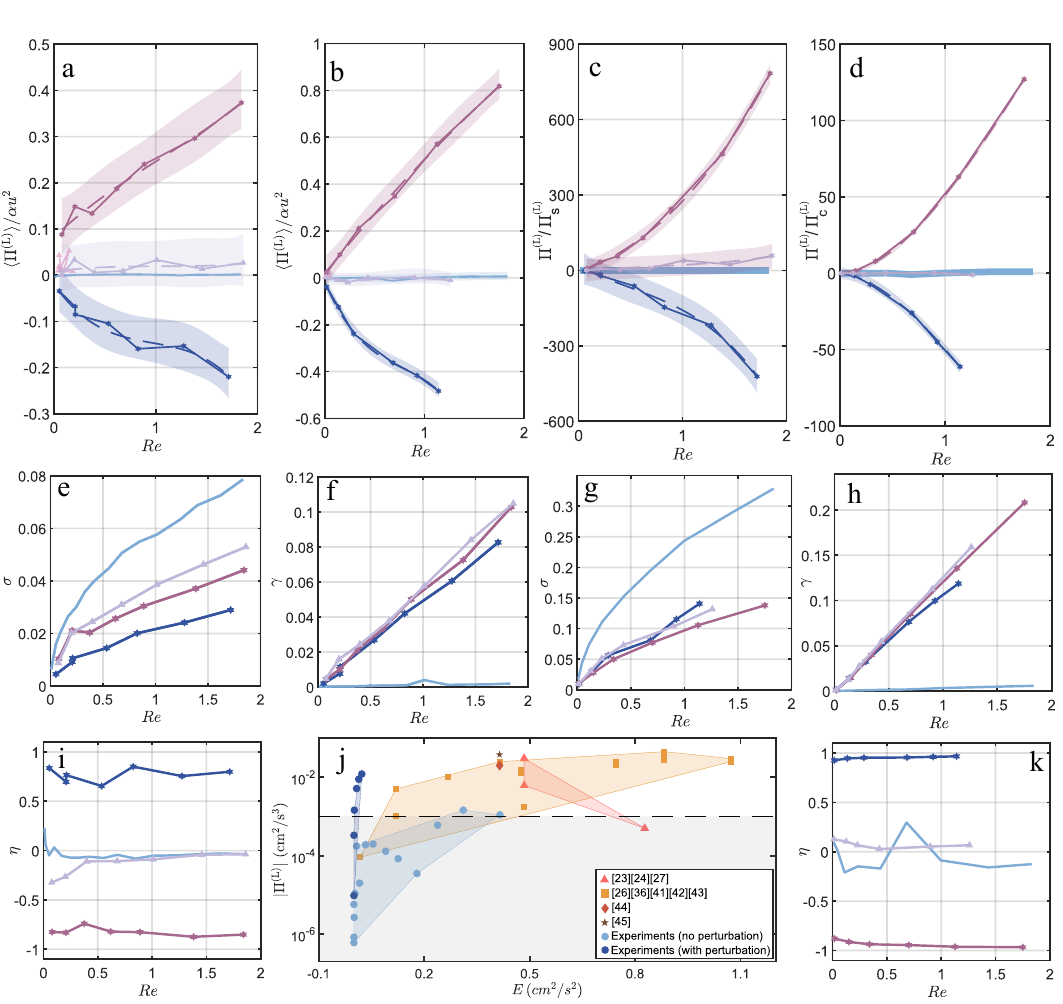}
    \caption{\textbf{Energy flux enhancement and element-wise understanding.} (\textbf{a} and \textbf{b}) $\Pi^{(\rm L)}$ for different Re with shear and cellular flows, respectively. (\textbf{c} and \textbf{d}) Energy flux ratio between enhanced and background flows at different Re with shear and cellular flow configurations. (\textbf{e} and \textbf{g}) Eigenvalue magnitudes of $s_{i,j}^{(L)}$ for shear and cellular flows, respectively. (\textbf{f} and \textbf{h}) Eigenvalue magnitudes of $\tau_{i,j}^{(L)}$ for shear and cellular flows, respectively. (\textbf{i} and \textbf{k}) Efficiency ($\eta$) for shear and cellular flows, respectively. (\textbf{j}) Magnitudes of $\Pi^{(\rm L)}$ versus kinetic energy contained in the system for background flow, perturbed flow and weak 2D turbulent flows in the literature\cite{fang2017multiple,fang2016advection,fang2021spectral,liao2013spatial,liao2014geometry,liao2012forcing, liao2015longrange, liao2015correlations,kelley2011spatiotemporal, ni2014extracting}. Flows with magnitudes of $\Pi^{(\rm L)}$ above the dashed line are typically considered as weak turbulence. 
    Panels \textbf{a}-\textbf{h}, \textbf{i}, and \textbf{k} share the same legend as Fig. \ref{fig2}. 
    All results are based on a cut-off length scale of 0.7. Shaded areas in \textbf{a}-\textbf{d} indicate 99\% confidence interval.}
    \label{fig4}
\end{figure}

In the following, we compare the magnitude of energy flux enhancement observed near a Re of unity, and systematically examine each constituent quantity in the energy flux expression (Eqn. \ref{eqn: tensor geometry})—$\gamma$, $\sigma$ and $\eta$—for both manipulated and background flows. Figure~\ref{fig4}a and b illustrate the enhancement of spectral energy flux across a range of Re for both shear and cellular background flows. Due to the nature of the physical perturbations, as a weak additional power input that changes Re, it is not feasible to generate perturbed and unperturbed flows with exactly matching Re. Thus, we perform a piecewise fit of the energy flux as a function of Re to capture the trend. 
To conservatively quantify enhancement, we compute the ratio of energy flux in the manipulated flow to that of the unperturbed background flow at the highest available Re (Fig.~\ref{fig4}c and d). When Re is near unity, we observe a spectral energy flux strength enhancement up to 800 times. This enhancement is evident in both the \(\theta \approx \pi/2\) and \(\theta \approx 0\) configurations.

The magnitude of the spectral energy flux is governed by Eqn.~\ref{eqn: tensor geometry}. To understand the origin of flux enhancement, we analyze each constituent quantity for $\Pi^{(L)}$. For both shear and cellular background flows, we observe that $\sigma$ increases approximately linearly with Re (Fig. \ref{fig4}e and g). 
In all manipulated and purely perturbed cases, $\sigma$ is consistently smaller than that of the background flow. This observation is consistent with physical intuition: the rate of strain tensor is a large-scale quantity, and unperturbed background flow should have the strongest rate of strain. 
In contrast, all perturbed cases exhibit a significant increase in $\gamma$ relative to the background flow (Fig. \ref{fig4}f and h). This is because the stress tensor is a small-scale quantity. Since the background flow is laminar, even modest additions of small-scale perturbations can substantially amplify $\gamma$. Furthermore, we find that the efficiency of energy flux (\(\eta\)), as governed by the tensor alignment angle \(\theta^{(L)}\), is significantly enhanced in both \(\theta \approx 0\) and \(\theta \approx \pi/2\) configurations (Fig. \ref{fig4}i and k). These results are consistent with our theoretical framework and suggest that turbulence, traditionally associated with high-Re flows, can be actively constructed and maintained in nominally laminar systems via geometric control of tensor alignment.

Based on these observations, we conclude that the enhancement of spectral energy flux at $\mathrm{Re} \sim \mathcal{O}(1)$ arises from two primary mechanisms. First, the introduction of small-scale directional perturbations significantly increases \(\gamma\). Second, once the directional biased perturbation favor mechanical angles that promote energy flux (\(\theta \approx 0\) and \(\theta \approx \pi/2\)), they enhance \(\eta\) by aligning the stress and rate of strain tensors thereby improving energy flux efficiency (\(\theta^{(L)} \approx 0\) and \(\theta^{(L)} \approx \pi/2\)). As a consequence, $\eta$ increases from near zero to values approaching 1 and -1, respectively. Together, these two mechanisms enable a remarkable amplification of spectral energy flux, even in nominally laminar flows.

Fig. \ref{fig4}j presents the spectral energy flux $\Pi^{(\rm L)}$ against the total kinetic energy in the system. While the precise value of the friction coefficient $\alpha$ is often not reported in prior studies, precluding direct calculation of Re, the experimental apparatus used in the literature closely resembles our setup. Assuming a comparable $\alpha$ and adopting a consistent length scale $L_i$, the Re becomes effectively proportional to the root-mean-square velocity. We therefore use the kinetic energy per unit mass, E, as a practical proxy for Re. Our results show that in both shear and cellular configurations, the background flow follows a similar trend as literature data—namely, spectral energy flux increases with growing kinetic energy. Strikingly, in the perturbed cases, we observe a substantial enhancement of energy flux even at very low kinetic energy levels. The horizontal dashed line marks a commonly accepted threshold above which 2D flows are considered to exhibit weak turbulence. Our perturbed flows surpass this threshold despite their low kinetic energy content ($\mathrm{Re} \sim \mathcal{O}(1)$), indicating that they already manifest a key hallmark of turbulence—substantial energy transfer between different length scales—well before the inertial forces dominate.

\section{Conclusion}
Our results demonstrate that turbulence—a phenomenon long believed to require large inertial forces—can be sustained at $\mathrm{Re} \sim \mathcal{O}(1)$ through targeted manipulation of tensor geometry \cite{si2025manipulating}. By introducing directional small-scale perturbations that both enhance stress magnitude and promote alignment between stress and rate of strain tensors, we achieve spectral energy flux levels comparable to weak turbulence. These findings challenge classical assumptions about the laminar–turbulent boundary and introduce a new pathway for engineering turbulent transport in non-inertial-dominated systems \cite{reynolds1883xxix,reynolds1895iv,orszag1973statistical}. 

Beyond their fundamental significance, these results provide practical strategies for improving mixing and transport in microfluidic environments where conventional turbulence is unattainable \cite{stroock2002chaotic,stone2004engineering}. Similarly, in biological systems operating at low Re—such as flow perturbed by the planktonic organisms \cite{houghton2018vertically,si2024biologically}—transport is often constrained by viscous dominance, presenting challenges for efficient mixing and biochemical signaling. In such environments, the lack of inertial forces typically precludes turbulent mixing, leaving diffusion as the dominant mechanism—one that is fundamentally slow and limiting in terms of throughput and efficiency. Our framework offers a mechanism to induce and control multiscale energy transport through tensor alignment, without increasing the net inertial forcing. This enables effective mixing and enhanced scalar transport even in regimes where classical turbulence is absent.



\clearpage

\section{Methods}

\subsection{Experimental method}
\subsubsection{Electromagnetically driven thin-layer flow system }
The experimental system for generating quasi-2D flow consisted of an acrylic frame housing a pair of copper electrodes positioned on opposite sides of the apparatus, with a central tempered glass plate serving as a separator between the shallow electrolyte layer and an underlying array of cylindrical neodymium magnets. The main frame and central glass substrate had dimensions of $96.5 \times 83.8\ \text{cm}^2$ and $81.3 \times 81.3\ \text{cm}^2$, respectively. 

To minimize friction, the top surface of the glass was coated with a hydrophobic agent (Rain-X), while the bottom surface was covered with a light-absorbing blackout film to reduce optical reflections. Cylindrical magnets (grade N52) were placed beneath the glass in specific patterns to induce flows in desired directions. Each magnet had an outer diameter of 1.27\,cm and a thickness of 0.64\,cm, providing a maximum surface magnetic flux density of 1.5\,T.

A thin electrolyte layer (6\,mm thick) consisting of a 14\% NaCl solution by mass was deposited on top of the glass. The fluid properties were: density $\rho = 1.101\,\text{g/cm}^3$ and kinematic viscosity $\nu = 1.25 \times 10^{-2}\,\text{cm}^2/\text{s}$. A quasi-2D flow was generated by applying a direct current (DC) across the electrolyte layer, inducing a Lorentz force through its interaction with the magnetic field. The Re of the flow was modulated by tuning the DC current, and the flow remained effectively 2D across all experiments.

To apply small-scale stress to couple with the rate of strain in the background flow, a moving rod array was employed. The rod array consisted of a $4\times4$ grid of rods, each with a diameter of \SI{2}{\milli\meter} and a center-to-center spacing of 1.5 cm between adjacent rods. The rods were fabricated using a 3D printer and coated with matte black paint to minimize light reflection. The motion of the rod array was controlled by a programmable linear actuator, operating at velocities of $0.2$--$1.0$\,cm/s for the shear flow and $0.1$--$1.1$\,cm/s for the cellular flow. To maintain better alignment, the rod array was moved at an appropriate speed to introduce perturbations while ensuring the disturbance did not overwhelm the background flow or disrupt the well-organized rate-of-strain field. For all the experiments, we made the rod move at around 2.5 times the background root-mean-square velocities.
The array moved along directions that had angles of $0$, $\pi/4$, and $\pi/2$ with respect to the rate of strain tensor's extensional direction.

\subsubsection{Particle tracking velocimetry}
The flow field was visualized by seeding the fluid with green fluorescent polyethylene tracer particles (Cospheric), which had a density of 1.0125 g/cm$^3$ and diameters ranging from 106 to 125 $\mu$m. The Stokes number St $\ll 1$ (on the order of $10^{-3}$) ensured that the particles accurately followed the flow \cite{ouellette2008transport}. To prevent particle clustering caused by surface tension, a small amount of surfactant was added. The influence of surface tension on particle motion was negligible, as confirmed by the absence of tracer movement when no external driving force was applied.

Flow visualization was conducted using a machine vision camera (Basler, acA2040-90$\mu$m), which captured a \SI{20}{\centi\meter} $\times$ \SI{20}{\centi\meter} region at the center of the setup with a resolution of $1600 \times 1600$ pixels. Image sequences were recorded at a frame rate of \SI{60}{frames\per\second} after confirming that the flow field had reached a stable and uniform state.
Following the acquisition, the motion of tracer particles was extracted using a PTV algorithm. Approximately 20,000 particles were captured per frame. The combination of high seeding density and high frame rate enabled high-resolution measurements of the Eulerian velocity field with excellent spatiotemporal accuracy.
The resulting particle velocities were interpolated onto a regular Eulerian grid using cubic interpolation, with a grid spacing of $\Delta x = 15$ pixels (corresponding to \SI{0.19}{\centi\meter}).
All quantities—including velocity, energy flux, filtered strain, strain eigenvalues and eigenvectors, and the angle between eigenvectors—were computed based on the full flow field. 

\subsection{Orientation of the Rate of Strain Tensor and Perturbation-Induced Stress Tensor}
\subsubsection{Rate of strain tensor orientation for cellular flow}
In this paper, we applied our physical perturbations around the hyperbolic point located between four counter-rotating cellular flows. The neighborhood around the hyperbolic point exhibits a well-ordered rate of strain tensor. 
A hyperbolic point is a structurally fundamental feature in fluid dynamics, emerging where fluid elements converge along one principal axis and diverge along another. This localized saddle-like topology produces a highly organized rate of strain field characterized by a symmetric and traceless structure. At such points, the strain tensor takes the canonical form:

\begin{equation}\label{Eqn_Sij}
    s_{ij}^{(\rm L)} = 
    \begin{bmatrix}
        \sigma & 0 \\ 
        0 & -\sigma
    \end{bmatrix},
\end{equation}
where $\sigma$ denotes the local stretching rate, and the extensional and compressive axes are aligned with orthogonal eigenvectors. The well-organized rate of strain tensor geometries represent an opportunity for us to apply directionally biased physical perturbations to manipulate the flow. 

\subsubsection{Rate of strain tensor orientation for shear flow}
In a linear shear flow, the velocity gradient is a constant $\rm K$. The rate of strain tensor for a low-pass-filtered linear shear with a cutoff length scale $\rm L$ that is reasonably smaller than the domain size is then:
\begin{equation}\label{Eqn_Sij}
    s_{ij}^{(\rm L)} = 
    \begin{bmatrix}
        0 & \frac{1}{2} \rm K \\ 
        \frac{1}{2} \rm K & 0
    \end{bmatrix}.
\end{equation}
The rate of strain remains traceless and symmetric in accordance with the incompressibility condition. Its eigenvectors are orthogonal, with the extensional eigenvector oriented at an angle of
$\pi/4$ or $5\pi/4$ relative to the direction of maximum velocity gradient.

\subsubsection{Perturbation-induced small-scale stress}
Here, we demonstrate how small-scale stress can be engineered using directionally biased physical perturbations. For simplicity, we model the small-scale disturbance as a velocity vector ${b}_i$. In a flow field with a well-organized large-scale rate of strain tensor and sufficient scale separation, this perturbation can be approximated as the difference between the total velocity field and its large-scale (low-pass filtered) velocity field as ${b}_i \approx {u}_i - {u}_i^{(L)}$, where ${u}_i$ is the instantaneous velocity field and ${u}_i^{(L)}$ is the low-pass filtered velocity field with a cutoff length of $L$. 

Under Cartesian coordinates, we define the unit vector $\mathbf{r_b}$ as: 
\begin{align}\label{Eqn_Sij}
    \mathbf{b} = &
    \|\mathbf{b}\| \mathbf{r^b}.
\end{align}

Then, the stress tensor can be written as:

\begin{align}\label{Eqn_Sij}
    \tau_s^{\rm (L)} = & (b_ib_j)^{\rm (L)}=  
    \|\mathbf{b}\|^{2} \mathbf{r_b}\mathbf{r_b}^{T} \\
    = & \|\mathbf{b}\|^{2} \mathbf{r_b}\otimes\mathbf{r_b}.
\end{align}

We define a normal stress matrix $R$ as $R=\mathbf{r_b}\otimes\mathbf{r_b}$. This matrix encodes the orientation of the perturbation-induced stress relative to the Cartesian coordinate frame. Since the rate of strain tensor is traceless due to incompressibility, only the deviatoric part of $R$ contributes to energy flux. Thus, we focus on the deviatoric component of the normal stress matrix, given by: $ R^ {dev}=R-\frac{1}{n}\mathrm{tr}(R)I$

It is easy to show that, in any dimension n, the extensional eigenvector is parallel to the $\mathbf{r_b}$ with an eigenvalue of $\sigma_r=\frac{n-1}{n}$ and all other n-1 compressional eigenvectors are orthogonal to $\mathbf{r_b}$ with eigenvalues of $-\frac{1}{n}$.

\newpage

\bibliography{Turbulence_at_low_Re} 
\bibliographystyle{sciencemag}

\clearpage
\newpage


\clearpage 

%

%
%
%
%
%
%


\section*{Acknowledgments}
\paragraph*{Funding:}
This work is supported by the U.S. National Science Foundation under Grant No. \mbox{CMMI-2143807} (L.F.) and \mbox{CBET-2429374} (L.F.).
\paragraph*{Author contributions:}
L.F. conceived the original idea and supervised the project. Z.Y. and X.S. ran the
experiments. Z.Y analyzed the data. L.F. and Z.Y. wrote the paper.
\paragraph*{Competing interests:}
The authors declare no competing interests.
\paragraph*{Data and materials availability:}
Correspondence and requests for materials should be addressed to Lei Fang.


\subsection*{Supplementary materials}
Materials and Methods\\
Supplementary Text\\
Figs. S1 to S3\\


\newpage


\renewcommand{\thefigure}{S\arabic{figure}}
\renewcommand{\thetable}{S\arabic{table}}
\renewcommand{\theequation}{S\arabic{equation}}
\renewcommand{\thepage}{S\arabic{page}}
\setcounter{figure}{0}
\setcounter{table}{0}
\setcounter{equation}{0}
\setcounter{page}{1} 


\begin{center}
\section*{Supplementary Materials for\\ \scititle}

Ziyue~Yu$^{1}$,
Xinyu~Si$^{1}$,
Lei~Fang$^{1,2,3\ast}$\\
\small$^\ast$Corresponding author. Email: lei.fang@pitt.edu
\end{center}

\subsubsection*{This PDF file includes:}
Materials and Methods\\
Figures S1 to S3


\newpage

\subsection*{Governing Equations of Energy Flux Across Scales}
\subsubsection*{Filter space technique}

The filter space technique utilizes a spatial filtering approach to extract spatially localized, scale-to-scale energy flux information directly from measured velocity fields. This filtering operation is commonly formulated as a convolution integral. For instance, the filtered component of a velocity field can be expressed in the following general form:

\begin{equation}\label{Eqn_Fitering}
    u_i^{(\rm L)} (\mathbf{x}) \equiv \int G^{(\rm L)}(\mathbf{r},\mathbf{x}) u_i(\mathbf{x} - \mathbf{r})d\mathbf{r},
\end{equation}
where $G^{(\rm L)}$ denotes a filter kernel, which acts as a low-pass filter. Superscript L indicates the cut-off length scale. Our result is not sensitive to the specific nature of the filter kernel. A sharp spectral filter was adopted in this paper and was further smoothed using a Gaussian window to mitigate potential ringing artifacts.

The governing equation for energy flux across scales is derived from the filtered N-S equation
\begin{equation}\label{Eqn_2DNS}
\quad \frac{\partial u_i}{\partial x_i} = 0,
\end{equation}
\begin{equation}\label{Eqn_2DNS}
    \frac{\partial u_i}{\partial {\rm t}} + u_j \frac{\partial u_i}{\partial x_j} = -\frac{1}{\rho}\frac{\partial p}{\partial x_i} + \nu\frac{\partial^2 u_i}{\partial x_j \partial x_j},
\end{equation}
in which $u_i$ is the $i^{\text{th}}$ component of the velocity, $\rho$ is the density, $p$ is the pressure, and $\nu$ is the kinematic viscosity. After applying the filter, we obtain the evolution equation for the filtered velocity field $u_i^{(L)}$ as:

\begin{equation}\label{Eqn_Filtered_2DNS}
    \frac{\partial u_i^{(\rm L)}}{\partial {\rm t}} + u_j^{(\rm L)} \frac{\partial u_i^{(\rm L)}}{\partial x_j} = -\frac{1}{\rho} \frac{\partial p^{(\rm L)}}{\partial x_i} + \nu\frac{\partial^2 u_i^{(\rm L)}}{\partial x_j \partial x_j} - \frac{\partial \tau_{ij}^{(\rm L)}}{\partial x_j}.
\end{equation}

To derive the evolution equation for the scale-dependent kinetic energy, defined as $\rm E^{(\rm L)}=\frac{1}{2}u_i^{(\rm L)}u_i^{(\rm L)}$, we take the inner product of $u_i^{(\rm L)}$ with the filtered momentum equation. After algebraic manipulation and rearrangement, we obtain the governing equation for energy flux across scales:

\begin{equation}\label{Eqn_MotionEnergy}
    \frac{\partial {\rm E}^{(\rm L)}}{\partial {\rm t}} = - \frac{\partial J_i^{(\rm L)}}{\partial x_i} - \nu \frac{\partial u_i^{(\rm L)}}{\partial x_j}\frac{\partial u_j^{(\rm L)}}{\partial x_i} - \Pi^{(\rm L)},
\end{equation}
where $\Pi^{(\rm L)}=-\tau_{ij}^{(\rm L)}s_{ij}^{(\rm L)}$, where the filtered rate of strain tensor is defined as $s_{ij}^{(\rm L)} = \frac{1}{2}(\partial u_i^{(\rm L)}\partial x_j + \partial u_j^{(\rm L)}/\partial x_i)$. The term $\Pi^{(\rm L)}$ represents the energy flux across scale $\rm L$ (specifically the transfer of energy between scales larger than $\rm L$ and those smaller than $\rm L$). A positive $\Pi^{(\rm L)}$ indicates a forward energy flux, i.e., energy is transferred from larger to smaller scales. Conversely, a negative $\Pi^{(\rm L)}$ indicates an inverse energy flux, where energy flows from smaller to larger scales.

The other terms in the equation are defined as follows: $\rm E^{(\rm L)}=\frac{1}{2}u_i^{(\rm L)}u_i^{(\rm L)}$ represents filtered kinetic energy, $J_i^{(\rm L)}=[u_i^{\rm L}(-\tau_{ij}^{(\rm L)}-\delta_{ij}\frac{P^{(\rm L)}}{\rho}+2\nu s_{ij}^{(\rm L)})]$ denotes the spatial transport of filtered kinetic energy, $\nu \frac{\partial u_i^{(\rm L)}}{\partial x_j}\frac{\partial u_j^{(\rm L)}}{\partial x_i}$ corresponds to dissipation. All of these terms describe intra-scale processes and do not involve energy transfer between different scales. Therefore, our focus is placed on the energy flux term $\Pi^{(\rm L)}$, which captures energy exchange between different scales of motions.

\subsubsection*{Energy flux calculation and decomposition}
Not all components of the stress tensor contribute equally to the energy flux across scales. The stress tensor $\tau_{ij}^{(\rm L)}$ can be decomposed into three components: 

\begin{equation}\label{Stress_Decomposition}
    \tau_{ij}^{(\rm L)} = \tau_L^{(\rm L)} + \tau_C^{(\rm L)} + \tau_S^{(\rm L)},
\end{equation}
where the Leonard stress, $\tau_L^{(\rm L)} = {(u_i^{(\rm L)}u_j^{(\rm L)})}^{(\rm L)} - u_i^{(\rm L)}u_j^{(\rm L)}$, is a small-scale quantity constructed from interactions between two large-scale components; the cross stress, $\tau_C^{(\rm L)} = {(u_i^{(\rm L)}(u_j - u_j^{(\rm L)}))}^{(\rm L)} + {(u_j^{(\rm L)}(u_i - u_i^{(\rm L)}))}^{(\rm L)}$ is a large-scale quantity, representing the interaction between large-scale and small-scale components; and the subgrid-scale Reynolds stress, $\tau_S^{(\rm L)} = {((u_i - u_i^{(\rm L)})(u_j - u_j^{(\rm L)}))}^{(\rm L)}$) captures the main contribution from the interaction of two small-scale components.


Similarly, the energy flux can also be decomposed into three corresponding contributions, associated with the Leonard stress, cross stress, and subgrid-scale Reynolds stress: 
\begin{eqnarray}\label{Eqn_PiDecomp}
    \Pi^{(\rm L)} & = &\Pi_L^{(\rm L)} + \Pi_C^{(\rm L)} + \Pi_S^{(\rm L)}, \\
    \Pi_L^{(\rm L)} & = & - \tau_L^{(\rm L)}s_{ij}^{(\rm L)}, \\
    \Pi_C^{(\rm L)} & = & - \tau_C^{(\rm L)}s_{ij}^{(\rm L)}, \\
    \Pi_S^{(\rm L)} & = & - \tau_S^{(\rm L)}s_{ij}^{(\rm L)}.
\end{eqnarray}

However, both theoretical and experimental results confirm that only $\Pi_S^{(\rm L)}$ contributes significantly to the net inverse energy flux across scales.
The term $\Pi_L^{(\rm L)}$ represents energy exchange between large-scale modes that are retained in the filtered velocity field, and its contribution vanishes upon spatial averaging. The cross-scale flux $\Pi_C^{(\rm L)}$, primarily affects regions near the forcing scale and contributes only a minor net inverse flux. Therefore, in this paper, we use the subgrid-scale stress $\tau_S^{(\rm L)}$ to compute the energy flux $\Pi^{\rm (L)}$.

Although the Leonard and cross terms are theoretically expected to contribute negligibly to the spectral energy flux, their inclusion can introduce significant contamination in the computed spatially averaged flux, particularly near domain boundaries. This contamination arises from edge effects associated with the filtering procedure, where artificial data must be introduced to enable convolution operations near the boundaries. In this study, zero-padding was employed to extend the measured data beyond the physical domain. While the resulting padding error is relatively small compared to the amplitude of the small-scale fluctuations $(u_i - u_i^{(L)})$ generated by the moving rods, it becomes non-negligible when added to or multiplied by large-scale velocity components $u_i^{(L)}$. To mitigate this issue, we focused our analysis exclusively on the subgrid-scale flux term $\Pi^{(L)}_S$, and for clarity and brevity, we omitted the subscript throughout the main text and figures. For the data we presented in this paper, we used the part that is away from the boundary, so the impact of padding is minimized.

\subsection*{Tensor geometry}  
\subsubsection*{Reformatting the spectral energy flux}

Under the incompressibility condition, the rate of strain tensor $s_{ij}$ is traceless, a property preserved after spatial filtering. 
Moreover, since the rate of strain tensor is symmetric and purely deviatoric in incompressible flows, only the deviatoric component of the stress tensor needs to be considered. When expressed in the eigenframe of the stress tensor, $\tau_{ij}$ becomes a diagonal matrix. Letting the positive eigenvalue for the rate of strain tensor is $\sigma$ and that of the stress tensor is $\gamma$, and the angle between the eigenframes is $\theta^{(\rm L)}$, we have:
\begin{eqnarray}\label{Eqn_TG}
    -\Pi^{(\rm L)} & = & \tau_{ij}^{(\rm L)}s_{ij}^{(\rm L)} \nonumber \\
    & = & \mbox{Tr} \left[
    \left(\begin{array}{cc}
         \gamma & 0 \\
         0 & -\gamma
    \end{array}\right) 
    \left(\begin{array}{cc}
         \cos (\theta^{(\rm L)}) & -\sin(\theta^{(\rm L)})  \\
         \sin(\theta^{(\rm L)}) & \cos (\theta^{(\rm L)})
    \end{array} \right)
    \left(\begin{array}{cc}
         \sigma & 0 \\
         0 & -\sigma
    \end{array}\right)
    \left(\begin{array}{cc}
         \cos (\theta^{(\rm L)}) & \sin(\theta^{(\rm L)})  \\
         -\sin(\theta^{(\rm L)}) & \cos (\theta^{(\rm L)})
    \end{array} \right)
    \right] \nonumber \\
    & = & 2\gamma\sigma\cos(2\theta^{(\rm L)}).
\end{eqnarray}

In two-dimensional turbulence, the inter-scale energy flux—its direction, magnitude, and transfer efficiency—can be fully characterized by the eigenvalues of the rate of strain and stress tensors, along with the angle between their respective eigenframes. While analogous formulations exist in 3D turbulence, they are considerably more complex and less analytically tractable. By contrast, the 2D case admits a compact, physically interpretable form. This tractability makes it feasible to manipulate energy transfer via controlled perturbations imposed on the flow field.

Notably, the stress tensor encapsulates information from smaller scales, whereas the rate of strain tensor reflects properties associated with larger scales. Although it is generally difficult to alter large-scale flow structures without simultaneously disturbing smaller-scale features, the reverse is not necessarily true: small-scale modulation may influence energy flux without significantly disrupting the large-scale rate of strain field. From this perspective, inputting specific perturbation induces a stress field that can align with the structured rate of strain field, making energy flux enhancement possible.

\begin{figure}[p]
    \centering
    \includegraphics[width=\textwidth]{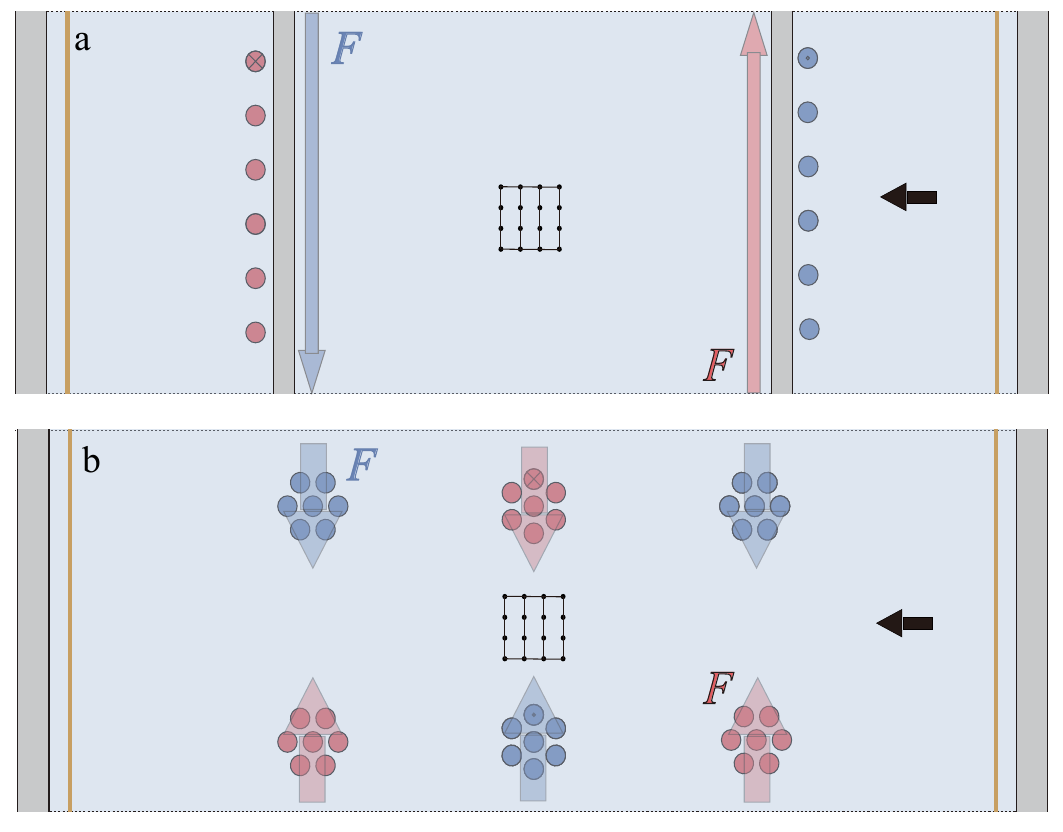}
    \caption{\textbf{Top views of experimental setup.}
(\textbf{a}) Schematic of shear flow. Two lines of permanent magnets are arranged with opposite polarities to generate a shear flow in the middle of the domain. The grey walls on both sides reinforce the no-slip boundary condition at the edges of the shear flow. The magnetic field driving the shear flow in the central region is a secondary magnetic field arising from the magnet arrays. A 4×4 rod array (black grid) is positioned at the center to introduce controlled perturbations. The black arrow indicates the direction of the applied DC current, while the blue and red arrows (F) denote the Lorentz forces induced in opposite directions.
(\textbf{b}) Schematic of cellular flow. Six groups of permanent magnets are arranged with alternating polarities to produce cellular flow patterns. As in the shear configuration, the black arrow indicates the direction of the DC current, and the blue and red arrows (F) represent the corresponding Lorentz forces.}
    \label{Ext.fig1}
\end{figure}

\clearpage

\begin{figure}[p]
    \centering
    \includegraphics[width=\textwidth]{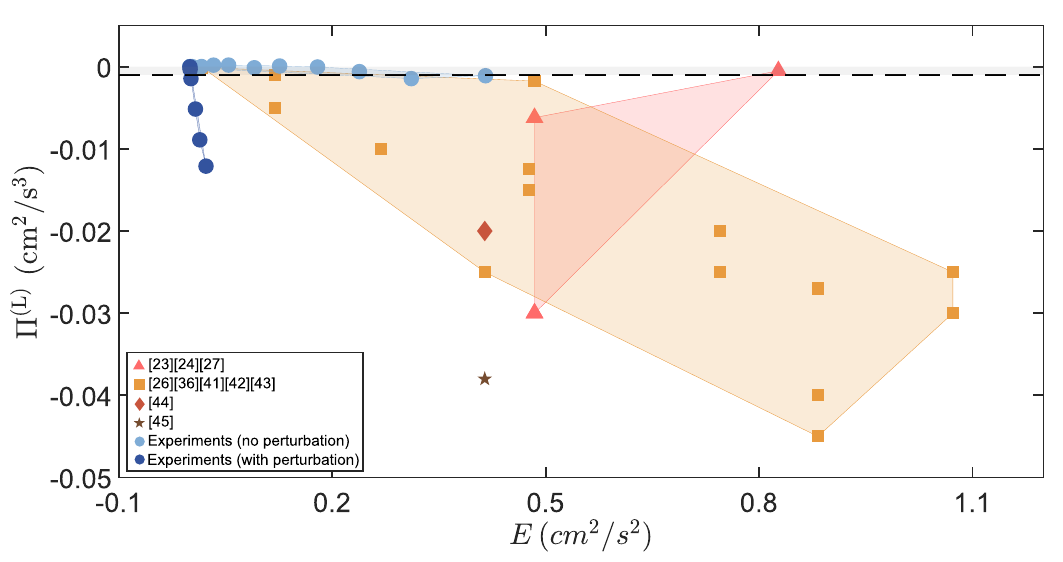}
    \caption{\textbf{Comparison of energy flux versus kinetic energy across literature and present experiments.} The spectral energy flux $\Pi^{(L)}$ is plotted against kinetic energy $E$ for various low-Reynolds-number turbulence experiments, including both prior studies and our enhancement experiments. Literature data points are grouped by symbol, with reference indices corresponding to those cited in the main text. In two-dimensional turbulence where energy cascades to larger scales, the spectral energy flux is negative. Notably, our enhanced cases exhibit flux magnitudes that exceed those reported in the literature, even at significantly lower kinetic energy levels. This demonstrates a novel pathway for turbulence-like energy transfer at Reynolds numbers near unity.}
    \label{Ext.fig1}
\end{figure}

\begin{figure}[p]
    \centering
    \includegraphics[width=\textwidth]{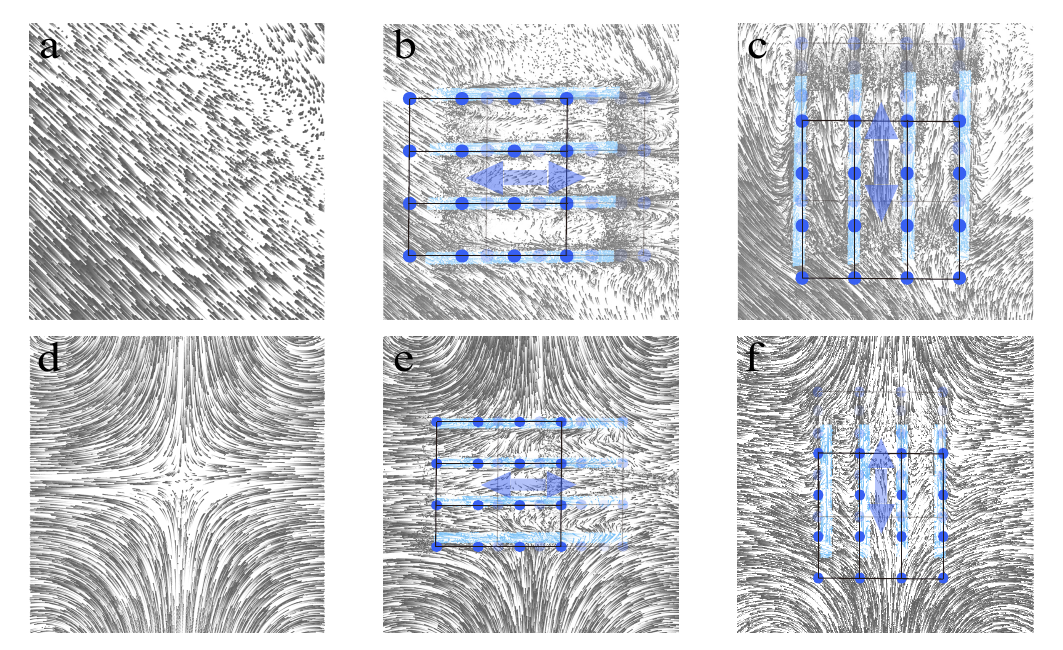}
    \caption{\textbf{Reconstructed time-lapse of particle motion} 
    Top row: Shear flow configuration. 
    (a) Unperturbed shear flow.
    (b) Perturbed shear flow with the perturbation direction aligned with the background rate of strain tensor’s extensional axis.
    (c) Perturbed shear flow with the perturbation direction perpendicular to the background rate of strain tensor’s extensional axis. 
    (d) Unperturbed cellular flow.
    (e) Perturbed cellular flow with the perturbation direction aligned with the background rate of strain tensor’s extensional axis.
    (f) Perturbed cellular flow with the perturbation direction perpendicular to the background rate of strain tensor’s extensional axis.  The Re are 1.23 (\textbf{a}), 1.27 (\textbf{b}), 1.38 (\textbf{c}), 1.43 (\textbf{d}), 1.14 (\textbf{e}), and 1.12 (\textbf{f}).}
        \label{Ext.fig1}
\end{figure}

\clearpage

\end{document}